\documentstyle[12pt,epsfig]{article}
\newcommand{\be}{\begin{equation}}
\newcommand{\ee}{\end{equation}}
\newcommand{\bea}{\begin{eqnarray}}
\newcommand{\eea}{\end{eqnarray}}

\newcommand{\bfk}{\mbox{\boldmath $k$}}

\newcommand{\bfp}{\mbox{\boldmath $p$}}
\newcommand{\bfP}{\mbox{\boldmath $P$}} 
\newcommand{\Lup}{\Lambda^\uparrow} 
 
\newcommand{\hup}{h^\uparrow} 
\newcommand{\hdown}{h^\downarrow} 
\newcommand{\nd}{\noindent}
\newcommand{\NP}[1]{{\it Nucl.\ Phys.}\ {\bf #1}}

\newcommand{\PL}[1]{{\it Phys.\ Lett.}\ {\bf #1}}
\newcommand{\PR}[1]{{\it Phys.\ Rev.}\ {\bf #1}}
\newcommand{\PRL}[1]{{\it Phys.\ Rev.\ Lett.}\ {\bf #1}}
\newcommand{\MPL}[1]{{\it Mod.\ Phys.\ Lett.}\ {\bf #1}}

\newcommand{\EPJ}[1]{{\it Eur.\ Phys.\ J.}\ {\bf #1}}
\newcommand{\IJMP}[1]{{\it Int.\ J.\ Mod.\ Phys.}\ {\bf #1}}

\begin{document}
\begin{flushright} 
DFTT 28/2001 \\ 
INFNCA-TH0108 \\ 
hep-ph/0109186 \\ 
\end{flushright} 
\vskip 1.5cm
\begin{center}
{\bf Transverse $\mbox{\boldmath$\Lambda$}$ polarization
 in semi-inclusive DIS}\\
\vskip 0.8cm
{\sf M.~Anselmino$^1$, D.~Boer$^2$, U.~D'Alesio$^3$, F.~Murgia$^3$}
\vskip 0.5cm
{\it $^1$ Dipartimento di Fisica Teorica, Universit\`a di Torino and \\
          INFN, Sezione di Torino, Via P. Giuria 1, I-10125 Torino, Italy}\\
\vspace{0.3cm}
{\it $^2$   Dept.\ of Physics and Astronomy, Vrije Universiteit Amsterdam, \\
De Boelelaan 1081, 1081 HV Amsterdam, The Netherlands} \\
\vspace{0.3cm}
{\it $^3$ INFN, Sezione di Cagliari and Dipartimento di Fisica,  
Universit\`a di Cagliari,\\
C.P. 170, I-09042 Monserrato (CA), Italy} \\
\end{center}

\vspace{1.5cm}

\begin{abstract}
Following a previous description of $\Lambda$ and $\bar\Lambda$
polarization in unpolarized $p\,$-$p$ interactions, within a 
perturbative QCD 
factorization scheme with new polarizing fragmentation functions, here we
investigate the transverse polarization of $\Lambda$'s and $\bar\Lambda$'s
produced in semi-inclusive DIS. Analytical expressions for both neutral and 
charged current exchange are given. Since quantitative predictions cannot be 
given at this stage and comparison with existing data are not yet significant,
we present the general formalism and a qualitative analysis displaying 
generic features of the $\Lambda$ and $\bar\Lambda$ polarization for specific 
scenarios. Different kinematical situations
are considered, corresponding to experiments currently able to study 
$\Lambda$ production in semi-inclusive DIS.
\end{abstract}

\vspace{0.6cm}

{}~~~PACS numbers: 13.88.+e, 13.60.-r, 13.15.+g, 13.85.Ni
\newpage 
\pagestyle{plain} 
\setcounter{page}{1} 
\nd 
{\bf 1. Introduction} 
\label{intro}
\vskip 6pt 
Transverse hyperon polarization in high energy, unpolarized hadron-hadron 
collisions has formed a long-standing challenge for theoretical models of 
hadronic reactions \cite{data,theo1}. The straightforward application of 
perturbative QCD and collinear factorization in the study of these 
observables is not successful, since the transverse polarization generated 
perturbatively in the partonic cross section would be proportional to 
$\alpha_s \, m_q/\sqrt{\hat{s}}$, which cannot be responsible for hyperon 
polarizations which may reach well over 20\%. 

Therefore, we have recently proposed a new approach~\cite{abdm01} to this 
problem based on perturbative QCD and its factorization theorems, and which 
includes polarization and intrinsic transverse momentum, $\bfk_\perp$, 
effects. It requires the introduction of a new type of leading twist 
fragmentation function (FF), one which is polarization and
$\bfk_\perp$-$\,$dependent, the so-called polarizing
FF \cite{Mulders-Tangerman-96,abdm01}. 

Ideally, our approach could be tested by first extracting these new functions
by fitting some available experimental data, and then using the same functions
to give consistent predictions for other processes. 
The problem with such a procedure is the actual availability of
experimental data in kinematical regions appropriate to the 
application of our scheme, which requires high center of mass energy and 
high transverse momentum $\bfp_{_T}$ of the produced hyperon. 

In Ref.~\cite{abdm01} we have used $p\,$-$p$ and $p\,$-$Be$ data in the range
$p_{_T} \simeq$ 1--3 GeV/$c$ (and $\sqrt s \simeq $ 25--60 GeV) to gain 
knowledge on the polarizing FF; we assumed this to be data resulting
from a current quark fragmentation, ignoring possible target fragmentation
mechanisms. However, this turns out to be a questionable assumption: 
a preliminary study of the unpolarized cross section -- at least in the 
kinematical region where both polarization and cross-section 
data are available (which is only a subset of the total region where data on
$\Lambda$ polarization have been published and used) -- shows that the 
formalism employed in our paper \cite{abdm01} results in at most a few 
percents of the experimental values. This casts doubts on the obtained 
polarizing FF, which, although they are reasonable in shape and magnitude 
and describe the $\Lambda$ and $\bar\Lambda$ polarization quite well,
should not be used to make predictions. All this is currently under
further investigation and a full assessment will be published elsewhere.  

In this paper we study transverse $\Lambda$ polarization, $P_{_T}^\Lambda$, 
in unpolarized semi-inclusive DIS, $\ell \, p\to \ell' \, \Lup \, X$. At 
present there exists only one small set of SIDIS data (from the NOMAD 
experiment \cite{nomad00,nomad01}) to compare with, but the kinematics and 
statistics are such that any comparison would be inconclusive. Other data 
will become available in the near future (HERMES, COMPASS). We develop a 
full consistent formalism to compute and evaluate $P_{_T}^\Lambda$, but,
according to the above considerations, instead of using the functions 
obtained from Ref.\ \cite{abdm01} to give predictions, we shall only use 
those results as a basis for selecting different models for the polarizing 
FF. This will be discussed extensively below.

A similar approach based on new polarization and $\bfk_\perp$-$\,$dependent
functions, has already been applied to the study of transverse
single spin asymmetries in inclusive particle production at large $x_{_F}$
and medium to large $\bfp_{_T}$~\cite{abm95}. In this case the 
new leading twist functions are called the Sivers distribution function
\cite{siv} and the Collins fragmentation function \cite{col}. Also here 
future data are awaited to further test the description of the asymmetries in 
terms of those functions. But also for pion or photon production, pQCD 
calculations for the unpolarized cross sections can be a factor 100 smaller
than data \cite{e706,ww,wang,zfpbl}, even in the central rapidity region and 
at large $\bfp_{_T}$ values; in those cases the discrepancy can
be explained by the introduction of $\bfk_\perp$ effects in the
{\em distribution\/} functions: these give a large, spin-independent,
enhancing factor, which brings the cross sections in agreement with
data. Such factors would not alter the calculation of the $\Lambda$
polarization, as they cancel in the ratio of cross sections. 
However, it is too early to draw a definite conclusion, and a more 
detailed study is in progress.

It should also be mentioned that there is an alternative approach based on 
perturbative QCD and its factorization theorems, namely the inclusion of 
higher twist functions of the type investigated by Qiu and Sterman 
\cite{QS-91} (see also \cite{ET-85} for an earlier, related
study, and \cite{boros} for a comprehensive review paper). 
These so-called soft gluon pole functions were applied to single spin
asymmetries in prompt photon production \cite{QS-91,Korotkiyan-T-94}; 
in the Drell-Yan process \cite{Hammon-97,BMT98,BQ01} and in pion production 
\cite{E-Korotkiyan-T-95,qs,KK00}. This method has also been applied to
transversely polarized $\Lambda$ production in electron-positron annihilation
\cite{Lu-Li-Hu} and recently, in hadron-hadron scattering \cite{KK01}. 
The latter study closely resembles our study of transversely polarized 
$\Lambda$ production using the polarizing FF and it is not excluded that 
there might be a connection between the two approaches (for indications, 
see \cite{BMT98,Ratcliffe-98}).

The process we are looking at is $\ell \, p \to \ell' \, \Lambda \, X$, where
the $\Lambda$ in general has a transverse momentum compared to the lepton
scattering plane. This transverse momentum is assumed to be much smaller
than the hard scale $Q$ and allows one to become sensitive to the
transverse momentum dependence of the fragmentation function. Strictly
speaking no factorization theorem has been proven for this particular
situation. But as explained in Ref. \cite{col} one expects a generalized
factorization theorem to hold here, similarly to the case of $e^+ e^-
\to h_1 \, h_2 \, X$ (in a sense the crossed channel), where hadrons $h_1$
and $h_2$ are in opposite back-to-back jets. These hadrons themselves are
nearly, but in general not exactly, back-to-back, giving rise to a
dependence on transverse momentum in the fragmentation functions.
For this process the factorization theorem has been proven rigorously
\cite{col2}. 

The main difference between this generalized factorization theorem and 
the usual collinear factorization theorems is the appearance
of so-called Sudakov factors. These factors appear naturally since fixed
order perturbation theory is not expected to give a proper description of
the cross section at small $p_T$. Resummation of large logarithms then
gives rise to these Sudakov factors (which are spin independent). In
practice, this resummation has the effect of broadening and lowering the
transverse momentum distribution and, as a consequence, 
transverse momentum dependent asymmetries \cite{dan}. 
At the moderate $Q^2$ values we are considering, the effect
of resumming such Sudakov logarithms is expected not to be important
(the numerical study in Ref. \cite{dan} supports this
expectation) and its effect on the average $k_\perp$ should be moderate
as well. Moreover, the Gaussian distributions employed in the present paper
can effectively describe the broadening of a resummed distribution (an
issue also discussed in Ref. \cite{col3}); therefore, we will not include 
such a refinement here. However, should future data require a higher average 
$k_\perp$ than a fit to $p\,$-$p$ data suggests, then such a resummation 
is most likely an important contribution. In the $p\,$-$p$ case \cite{abdm01} 
no large logarithm resummation aspects were considered, since the explicit 
$k_\perp$ distribution is less important (due to the large $p_T$), hence 
the use of a distribution highly peaked around the average $k_\perp$ suffices.

The outline of this paper is as follows. 
In Section 2 we will discuss the details of the polarizing fragmentation 
functions and recall our approach for $p\,p \to \Lup \, X$ in order to set 
the notation. In Section 3 we will elaborate on a Gaussian model for the
$\bfk_\perp$-$\,$dependence of the fragmentation functions and 
in Section 4 we present our study of various electroweak SIDIS processes, 
where analytical expressions for both neutral and charged current exchange 
are given (this is done more completely in the Appendix).

In Section 5 we present some numerical results; since quantitative 
predictions cannot be made at this stage and comparison with existing data 
is not yet significant, we select simple scenarios for the polarizing FF
and present a qualitative analysis displaying generic features of $\Lambda$ 
and $\bar\Lambda$ polarization. Different kinematical situations are 
considered, corresponding to experiments currently able to study $\Lambda$ 
production in semi-inclusive DIS, i.e.\ NOMAD, HERMES, COMPASS and E665.

\vskip 18pt 
\nd 
{\bf 2. Polarizing fragmentation functions}
\label{polFF}
\vskip 6pt

The main idea behind the polarizing FF is that a transversely polarized
hyperon can result from the fragmentation of an {\em unpolarized\/} quark,
as long as the hyperon has a nonzero transverse momentum compared to the quark
(otherwise it would violate rotational invariance). This effect
need not be suppressed by inverse powers of a large energy scale, 
such as the center of mass energy $\sqrt{s}$ in $p\,$-$p$ scattering or 
the momentum transfer $Q$ in SIDIS. Actually, in $p\,$-$p$ scattering it 
appears with an inverse power of the transverse momentum $p_{_T}$ 
of the hyperon, which should be large enough for factorization to hold, but 
in practice is still much smaller than $\sqrt{s}$. 

In order to be able to apply a factorization theorem to the process $p\,p 
\to \Lambda \, X$, one has to require that both $\sqrt{s}$ (the overall 
c.m.\ energy) and $p_{_T}$ (the magnitude of the transverse momentum of the 
detected hadron in the $p\,$-$p$ c.m.\ frame) are large. Such a configuration 
originates from a large momentum transfer in the partonic interactions, so 
that perturbative calculations are justified.
In semi-inclusive DIS processes, $\ell \, p \to \ell' \, \Lambda \, X$, one 
only needs to require the momentum transfer $Q$ to be large. 
Our results will be given in the virtual photon (or vector boson)-$p$
c.m.\ frame, and $\bfp_{_T}$ is measured relatively to this direction;  
neglecting intrinsic motion in the incoming proton, in this frame 
$\bfp_{_T}$ coincides with the hadron intrinsic $\bfk_\perp$,
so that it can be small and one becomes sensitive to the precise 
$\bfk_{\perp}$ dependence of the polarizing FF. Therefore, in contrast to
Ref.\ \cite{abdm01}, where the $\bfp_{_T}$ dependence of the observable was
sufficiently described by evaluation of the polarizing FF 
at an {\it effective}, average  $k_\perp^{\,0}(z)$, here we will assume an 
explicit Gaussian $k_\perp$ dependence of the polarizing FF (cf.\ the next
Section for details). This will allow 
us to investigate $\ell \, p \to \ell' \, \Lup \, X$ for small 
(and in principle, after 
including hard perturbative corrections, also large) $p_{_T}$. An important 
point is that one should restrict to the current
fragmentation region, i.e.\ $x_{_F} > 0$ 
(many models for the target fragmentation region exist,
but we cannot compare our approach to those models, because they are simply 
not addressing the high energy QCD aspect of polarized $\Lambda$ production). 

In order to set the notation we recall that in our approach for the 
$p\,p \to \Lup \, X$ case, the transverse hyperon polarization in 
unpolarized hadronic reactions at large $p_{_T}$ can be 
written as follows~\cite{abdm01}
\begin{eqnarray}
 &\,\,\,\,P_{_T}^{\,\Lambda}(x_{_F},p_{_T})&\, = \>\>
  \frac{d\sigma^{pp\to\,\Lambda^\uparrow\,X}-
   d\sigma^{pp\to\,\Lambda^\downarrow\,X}}
  {d\sigma^{pp\to\,\Lambda^\uparrow\,X}+
   d\sigma^{pp\to\,\Lambda^\downarrow\,X}} \label{ptlh}\\
 &\!\!\!\!\!\!\!\!\!\!\!\!\!\!\!\!\!\!\!\!\!\!\!\!=\!\!\!\!\!\!&
  \!\!\!\!\!\!\!\!\!\!\!\!\!\!\!\!\!\!\!\!\!
 \frac{\sum\,\int dx_a\,dx_b \int\! d^2\bfk_{\perp}\,
 f_{a/p}(x_a)\, f_{b/p}(x_b)\, d\hat\sigma(x_a,x_b;\bfk_{\perp})\,
 \Delta^{\!N}\!D_{\Lup\!/c}(z,\bfk_{\perp})}
 {\sum\,\int dx_a\,dx_b \int\! d^2\bfk_{\perp}\,
 f_{a/p}(x_a)\, f_{b/p}(x_b)\, d\hat\sigma(x_a,x_b;\bfk_{\perp})\,
 \hat D_{\Lambda/c}(z,\bfk_{\perp})}\,,\nonumber
\end{eqnarray}
where $f_{a/p}(x_a)$ and $f_{b/p}(x_b)$ are the usual unpolarized parton 
densities; $d\hat\sigma(x_a,x_b;\bfk_{\perp})$ 
describes the lowest order partonic cross section with the inclusion 
of $\bfk_{\perp}$ effects; the $\sum$
takes into account all possible elementary interactions;    
$\hat D_{\Lambda/c}(z,\bfk_{\perp})$ and
$\Delta^{\!N}\!D_{\Lup\!/c}(z,\bfk_{\perp})$
are respectively the unpolarized and the polarizing 
FF \cite{Mulders-Tangerman-96,abdm01} for the process $c\to\Lambda+X$.
The polarizing FF is defined as: 
\bea 
\Delta^{\!N}\!D_{\hup\!/a}(z, \bfk_{\perp}) &\equiv& 
\hat D_{\hup\!/a}(z,\bfk_{\perp}) - \hat D_{\hdown\!/a}(z,\bfk_{\perp})  
\label{deld1}\\ 
&=& \hat D_{\hup\!/a}(z,\bfk_{\perp})-\hat D_{\hup\!/a}(z,-\bfk_{\perp}) \>, 
\nonumber 
\eea 
and denotes the difference between the density numbers  
$\hat D_{\hup\!/a}(z, \bfk_{\perp})$ and  
$\hat D_{\hdown\!/a}(z,$ $\bfk_{\perp})$ 
of spin 1/2 hadrons $h$, with longitudinal momentum fraction $z$, transverse  
momentum $\bfk_{\perp}$ and transverse polarization $\uparrow$ or  
$\downarrow$, inside a jet originated from the fragmentation of an  
unpolarized parton $a$. From the above definition it is clear that the 
$\bfk_{\perp}$ integral of the function vanishes and that, due to parity
invariance, the function itself vanishes in case the transverse momentum 
and transverse spin are parallel. Conversely, one can write
\be  
\hat D_{\hup\!/q}(z, \bfk_\perp) = \frac 12 \> \hat D_{h/q}(z, k_\perp) +  
\frac 12 \> \Delta^{\!N}\!D_{\hup\!/q}(z, k_\perp) \>  
\frac{\hat{\bfP}_h \cdot (\bfp_q \times \bfk_\perp)} 
{|\bfp_q \times \bfk_\perp|} \label{lamfn}
\ee
for an unpolarized quark with momentum $\bfp_q$ which fragments into  
a spin 1/2 hadron $h$ with momentum $\bfp_h = z \bfp_q + \bfk_\perp$
($\bfp_q \cdot \bfk_\perp = 0$) and polarization vector along the 
$\uparrow \> = \hat{\bfP}_h$ direction; 
$\hat D_{h/q}(z, k_\perp) = \hat D_{\hup\!/q}(z, \bfk_\perp) +   
\hat D_{\hdown\!/q}(z, \bfk_\perp)$ is the $k_\perp$-$\,$dependent 
unpolarized fragmentation function, where $k_\perp = |\bfk_\perp|$.
The variable $z$, in the kinematical region considered in Ref.\ \cite{abdm01},
is very close to the light-cone variable $\xi = p_h^+/p_q^+$. 

Throughout the paper we will adopt also the following notations:
\be
\Delta^{\!N}\!D_{\hup\!/q}(z, \bfk_\perp) \equiv 
\Delta^{\!N}\!D_{\hup\!/q}(z, k_\perp) \> 
\frac{\hat{\bfP}_h \cdot (\bfp_q \times \bfk_\perp)} 
{|\bfp_q \times \bfk_\perp|} = 
\Delta^{\!N}\!D_{\hup\!/q}(z, k_\perp) \> \sin\phi \>, 
\label{lamfn2}
\ee
where $\phi$ is the angle between $\bfk_\perp$ and $\bfP_h$, which,
in our configuration (as explained later), is the difference between
the azimuthal angles of $\bfP_h$ and $\bfk_\perp$, 
$\phi = \phi_{P_h} - \phi_{k_\perp}$.

Eq.~(\ref{ptlh}) is based on some simplifying assumptions (for a detailed
discussion we refer to \cite{abdm01}): 
(1)~The $\Lambda$ polarization is assumed to be generated in the
fragmentation process; 
(2)~The intrinsic $\mbox{\boldmath$k$}_\perp$ effects in the unpolarized 
initial nucleons are neglected; 
(3)~The $\Lambda$ FF's also include $\Lambda$'s coming from decays of 
other hyperon resonances. We will make these same assumptions in the present
SIDIS study. 

Determining the overall
sign of the $\Lambda$ polarization is important, hence we will specify the
frame of choice whenever appropriate. In this respect, we would like to add
this information to Eq.\ (4) of Ref.\ \cite{abdm01}, which gives the
relation between the function $D_{1T}^\perp$ defined in 
\cite{Mulders-Tangerman-96} and $\Delta^{\!N}\!D_{\hup\!/a}$ as will be 
used in the present paper;
in terms of the angle $\phi$ defined in Eq.\ (\ref{lamfn2}) the exact 
relation is:
\be 
\Delta^{\!N}\!D_{\hup\!/a}(z, \bfk_{\perp})   
= - 2 \, \frac{k_\perp}{z M_h} \; \sin\phi \;  
D_{1T}^\perp(z, k_\perp) \>. 
\label{rel}  
\ee 

We would also like to note that since the polarizing FF are chiral-even 
functions, as opposed to the Collins FF for instance, the observable of
interest here 
-- transverse $\Lambda$ polarization in SIDIS -- is independent of the lepton 
scattering plane. In other words, the $\Lambda$ polarization does not average 
to zero when one integrates over all directions of the back-scattered lepton. 
This fact allows one to study the neutral current process $\nu \, p \to 
\nu \, \Lup \, X$. Also, because we are dealing with chiral-even functions, 
charged current exchange processes -- which select fixed helicities -- may
give access to them, in contrast to the case of chiral-odd functions. 

\vskip 18pt 
\goodbreak
\nd 
{\bf 3. A Gaussian model for $\bfk_\perp$-$\,$dependent
fragmentation functions}
\label{gauss}
\nobreak
\vskip 6pt

The goal of this paper is to give a well defined formalism to compute 
the transverse $\Lambda$, $\bar{\Lambda}$ polarization in unpolarized 
semi-inclusive DIS and to present some qualitative and generic numerical
results.

As a first approach, we will take into account only leading twist
(in the $1/Q$ power expansion) and leading order (in the coupling
constant power expansion) contributions, looking at the process in the
virtual boson-target nucleon c.m. reference frame (VN frame).
Under these conditions the elementary virtual boson-quark scattering
is collinear (and along the direction of motion of the virtual boson)
and the transverse momentum of the final hadron with respect 
to the fragmenting quark, $\bfk_\perp$, coincides with the hadron
transverse momentum, $\bfp_{_T}$, as measured in the VN frame.
Therefore, to study the $\ell\,p\to\ell'\,\Lup\,X$
process and its dependence on the observed hadronic variables,
a complete knowledge of the $\bfk_\perp$ dependence of the (unpolarized 
and polarizing) fragmentation functions is required.    
To this end we will consider a Gaussian model for the explicit
$\bfk_\perp$ dependence of our FF.
By imposing proper constraints on the polarizing fragmentation 
function $\Delta^{\!N}\!D_{\Lup\!/q}(z,\bfk_\perp)$, 
we get an explicit Gaussian model for the FF, which enables us to investigate 
several interesting processes in the case of SIDIS. 
This procedure is described in detail in what follows.

For simplicity, we assume that the $\bfk_\perp$ dependence 
of the FF is the same for all flavours of the fragmenting quark,
a choice which remains to be tested but that seems quite reasonable.

We then write the unpolarized and polarizing FF for the $q\to\Lambda+X$
process in the following general form:
\bea
 \hat{D}_{\Lambda/q}(z,\bfk_\perp) &=&
 \hat{D}_{\Lambda/q}(z,k_\perp) \> = \> 
 \frac{d(z)}{M^2}\,\exp\Bigl[\,-\frac{k_{\perp}^{\!\ 2}}
 {M^2 f(z)}\,\Bigr] \, ,
\label{defd} \\
 \Delta^{\!N}\!D_{\Lup\!/q}(z,k_\perp) &=&
 \frac{\delta(z)}{M^2}\,\frac{k_\perp}{M}\,
 \exp\Bigl[\,-\frac{k_{\perp}^{\!\ 2}}
 {M^2 \varphi(z)}\,\Bigr]\, ,
\label{defded}
\eea
where $M=1$ GeV/$c$ is a typical hadronic momentum scale and $f(z)$,
$\varphi(z)$ are generic functions of $z$, which we choose to be of
the form $Nz^a(1-z)^b$. This simple behaviour naturally allows to take 
into account some general features of the $k_\perp$ dependence of the FF;
the fast exponential decrease allows to formally extend the integration 
region to the full range $0<k_\perp<+\infty$ (rather than to a typical
intrinsic $k_\perp$ range) making all analytical computations much easier,
with negligible numerical differences.   

Eqs.\ (\ref{defd}) and (\ref{defded}) must satisfy the positivity bound
\be
\frac{|\Delta^{\!N}\!D_{\Lup\!/q}(z,k_\perp)|}
{\hat{D}_{\Lambda/q}(z,k_\perp)} =
\frac{|\delta(z)|}{d(z)}\,\frac{k_\perp}{M}\,
\exp\Biggl[\,-\frac{k_\perp^{\!\ 2}}{M^2} \left( 
\frac{1}{\varphi}-\frac{1}{f} \right) \,\Biggr] \leq 1\>,
\label{posi}
\ee    
which, with  $\varphi(z) = r f(z)$, implies $r<1$ and 
\be
\label{pos2}
\frac{|\delta(z)|}{d(z)}  \leq   \left
[\frac{2\,e}{f(z)}\frac{1-r}{r}\right]^{1/2}
\>, 
\ee
in order for Eq.\ (\ref{posi}) to hold true for each value of $z$ and 
$k_\perp$.

It is easily verified that the functions $d(z)$, $f(z)$ in
Eq.~(\ref{defd}) are simply related to the usual, unpolarized and 
$\bfk_\perp$-$\,$integrated FF and to the hadron mean
squared transverse momentum inside the observed fragmentation jet,
$\langle k_\perp^{\!\ 2}(z)\rangle$: 
\bea
 D_{\Lambda/q}(z) &=& \int\! d^2\mbox{\boldmath$k$}_\perp\,
 \hat{D}_{\Lambda/q}(z,k_\perp) \> = \> \pi\,d(z)\,f(z)\, ,
\label{dz}\\
 \langle k_\perp^{\!\ 2}(z) \rangle &=& \frac{\int d^2\bfk_\perp\,
 k_{\perp}^{\!\ 2}\, \hat{D}_{\Lambda/q}(z,k_\perp)}
 {\int d^2\bfk_\perp\,\hat{D}_{\Lambda/q}(z,k_\perp)} \> = \> M^2\,f(z)\, ,
 \label{fz}
\eea
so that
\be
 d(z) = M^2\,\frac{D_{\Lambda/q}(z)}
 {\pi\,\langle  k_\perp^{\!\ 2}(z)\rangle} \>, \quad\quad\quad
 f(z) = \frac{\langle k_\perp^{\!\ 2}(z)\rangle}{M^2}\, \cdot
\label{fdz}
\ee

At present some experimental information on
$\langle k_\perp^{\!\ 2}(z)\rangle$ is available for pions but not yet for   
$\Lambda$ particles. However, several experiments, in different
kinematical configurations, are studying or plan to study $\Lambda$
production in SIDIS and could give information on this observable.

To obey Eq.\ (\ref{pos2}) in a most natural and simple way we can write
\be
\frac{\delta(z)}{M^3} = \left[N_q \,\frac{z^\alpha (1-z)^\beta} 
{\alpha^\alpha \beta^\beta/(\alpha+\beta)^{(\alpha+\beta)}}\right]\,
\frac{D_{\Lambda/q}(z)}{\pi[\langle k_\perp^{\!\ 2}(z)\rangle]^{3/2}}\,
[2e(1-r)/r]^{1/2} \label{pardel} 
\ee
with $\alpha, \beta > 0$, $|N_q|\le 1$.

In this approach $\delta(z)$ is an unknown function depending on the  
parameters $\alpha$, $\beta$, $N_q$ and $r$, while $\varphi(z)$ is fixed by 
$r$, if we assume to know the functions $\langle k_\perp^{\!\ 2}(z)\rangle$
and $D_{\Lambda/q}$.

If one were to demand consistency with the results of Ref.\ \cite{abdm01}
$\delta(z)$ and $\varphi(z)$ could be fixed, 
allowing one to give predictions for $P^\Lambda_{_T}$ in SIDIS. 
However, in the introduction we have already expressed the problems of such 
an approach and we will not pursue this here. 
 
Instead we will present the analytical formalism, which is meaningful and 
valid in the appropriate regions, and show some numerical results 
for different choices of $N_q, \alpha, \beta, r$ based on some of
the qualitative features obtained from our earlier analysis of hadronic data 
and where we choose the expression of $\langle k_\perp^{\!\ 2}(z)\rangle$ 
valid for pions at LEP energies. 

Collecting all results, we can finally give 
explicit $z$ and $\bfk_\perp$-$\,$dependent expressions for the
unpolarized and polarizing FF which we will use to investigate the
consequences of the analytical expressions given in Section 4: 
\be
\hat{D}_{\Lambda/q}(z,k_\perp) =
\frac{D_{\Lambda/q}(z)}{\pi\,\langle k_\perp^{\!\ 2}(z)\rangle}\,
\exp\Bigl[\,-\frac{k_\perp^{\!\ 2}}
{\langle k_\perp^{\!\ 2}(z)\rangle}\,\Bigr]\,,
\label{dfin}
\ee
\be
\Delta^{\!N}\!D_{\Lup\!/q}(z,k_\perp) =
\frac{\delta(z)}{M^2}\,\frac{k_\perp}{M}\,
\exp\Bigl[\,-\frac{k_{\perp}^{\!\ 2}}
{r \, \langle k_\perp^{\!\ 2}(z)\rangle} \,\Bigr] \,,
\label{dedfin}
\ee
with $\delta(z)$ taken as in Eq.\ (\ref{pardel}).

\vspace{18pt}
\goodbreak
\nd
{\bf 4. Analytical results}
\nobreak
\vspace{6pt}
\nobreak

In this Section we present the analytical expressions -- in terms of the
polarizing FF -- for the transverse $\Lambda$, $\bar\Lambda$ polarization 
in semi-inclusive DIS, both for neutral and charged current interaction.
Electroweak interference effects will be neglected, 
since they are hardly accessible by present or near-future experiments, 
at least in this context. We will apply the obtained 
expressions using different scenarios for the polarizing FF, but we expect
them to be useful in the future, when either relevant SIDIS data are  
available or the polarizing FF have been extracted from other processes. 

We give the expression of the polarization as a function of
the usual DIS variables, $x=Q^2/2p\cdot{q}$, 
$y=p \cdot q/p\cdot \ell$ and $z_h = p\cdot p_h/p \cdot q$,
where $\ell$, $p$, $q$, $p_h$ are the 4-momenta of the initial lepton, 
the proton, the virtual boson and the observed hadron respectively. 
Notice that for SIDIS processes in the VN frame $z_h$ coincides
with the light-cone variable $\xi = p_h^+/p_q^+$, where $p_q$ is the 
four-momentum of the fragmenting quark, $p_q = (W/2)(1,0,0,1)$ and
$W = [(1-x)\,ys]^{1/2}$ is the total energy in the VN frame. 

When appropriate, the observables will be averaged over the kinematical
range of the variables effectively covered by the particular
experimental situation considered.

As already discussed, we will present our results in the virtual 
boson-proton c.m. reference frame (VN frame). More specifically,
we take the $\hat{z}$-axis of our frame along the direction of motion 
of the exchanged virtual boson; the $\hat{x}$-axis is chosen along the 
transverse momentum $\mbox{\boldmath$p$}_{_T}$
of the observed hadron with respect to the $\hat{z}$-axis.
Since at leading twist and leading order the 
fragmenting quark moves collinearly to the virtual boson,
$\bfp_{_T}$ coincides with the hadron transverse momentum
with respect to the fragmenting quark, $\bfk_\perp$.
In this configuration, the transverse (with respect to the hadron
production plane) polarization is measured along the $+\hat{y}$-axis,
and the angle between the hadron intrinsic transverse momentum and
its polarization is fixed, $\phi=\pi/2$. For further clarity we show
in Fig. 1 our kinematical configuration: the up and down arrows show 
explicitly the direction of {\it positive} polarization for 
$\Lambda$'s produced in the four quadrants.  

The transverse $\Lambda$ polarization is given by
\be
P_{_T}^{\,\Lambda}(x,y,z_h,p_{_T})=
\frac{d\sigma^{\Lambda^\uparrow}-d\sigma^{\Lambda^\downarrow}}
{d\sigma^{\Lambda^\uparrow}+d\sigma^{\Lambda^\downarrow}} \>,
\label{updow}
\ee
with
\be
d\sigma^{\Lambda^{\uparrow(\downarrow)}}=
\frac{d\sigma^{\ell p\to\ell' \Lambda^{\uparrow(\downarrow)} X}}
{dx\,dy\,dz_h\,d^2\bfp_{_T}}=\sum_{i,j}\,f_{q_i/p}(x)\,
\frac{d\hat{\sigma}^{\ell q_i\to\ell' q_j}}{dy}
\hat{D}_{\Lambda^{\uparrow(\downarrow)}\!/q_j}(z_h,p_{_T}) \>,
\label{dgen}
\ee
where $\hat{D}_{\Lambda^{\uparrow(\downarrow)}\!/q}$ is defined by 
Eqs. (\ref{lamfn}) and (\ref{lamfn2}), with $\sin\phi=+(-)\,1$;
$i$,$j$ indicate different quark flavours and the sum includes
both quark and antiquark contributions; $(\ell,\ell')$ stands for
$(\ell^{\mp},\ell^{\mp})$, $(\nu,\nu)$, $(\bar\nu,\bar\nu)$
(NC contributions) and $(\nu,\ell^-)$, $(\bar{\nu},\ell^+)$, $(\ell^-,\nu)$,
$(\ell^+,\bar\nu)$ (CC contributions); in the NC case the partonic 
cross sections $d\hat{\sigma}^{\ell q_i\to\ell' q_j}/dy$ include a 
$\delta_{ij}$ term. 

Analytical expressions for the polarization can be derived from 
Eqs.\ (\ref{updow}) and (\ref{dgen}), by inserting the explicit formulae
of the elementary cross sections; the results are formally very similar 
to the ones presented in the case of longitudinal hadron polarization
(with unpolarized initial target) in Ref.~\cite{epjc01}.
Clearly, some appropriate modifications have to be performed
in order to adapt those results to the case of transverse polarization,
with the inclusion of $\bfk_\perp$ effects.
In particular, $\bfk_\perp$-$\,$integrated and longitudinally polarized FF, 
$\Delta{D}_{\Lambda/q}(z_h)$, must be substituted by the corresponding
polarizing FF, $\Delta^{\!N}\!D_{\Lup\!/q}(z_h,k_\perp)$. 
Furthermore, the sign of some of the terms in the numerator of
the polarization changes from negative to positive, a fact which can be
understood by a careful inspection of the formulae derived in
Ref.~\cite{epjc01}. The complete formulae can also be obtained from 
Ref.~\cite{bjm} by expressing the function $D_{1T}^\perp$ in terms
of $\Delta^{\!N}\!D_{\Lup\!/q}$ using Eq.\ (\ref{rel}).

We list the full results for the transverse $\Lambda$ and $\bar\Lambda$
polarization, only neglecting electroweak interference effects, 
in the Appendix. In some cases the full expressions 
are quite involved and not too informative. In this Section we present and 
discuss, for the processes of phenomenological interest, approximate
expressions which are valid with high accuracy for $\Lambda$ particles
and are much simpler. These expressions are 
obtained by neglecting terms containing nonleading quark contributions
both in the partonic distributions and fragmentation functions.
Isospin symmetry is assumed to hold, that is we
take $(\Delta^{\!N})D_{\Lambda/d} = (\Delta^{\!N})D_{\Lambda/u}$.

A similar argument holds also for the $\bar\Lambda$ case, at least in 
the region $z_h > 0.2$, where nonleading (or sea) quark contributions to 
the FF are relatively small. However, since the $\bar\Lambda$ case always 
involves at the same time contributions of the type
$q\,(\Delta^{\!N})D_{\Lambda/\bar{q}}$ and
$\bar{q}\,(\Delta^{\!N})D_{\Lambda/q}$, with combined sea and valence 
contributions, these approximations should be taken with more care,
also depending on the different ranges of the SIDIS variables
$x$, $y$, and $z_h$.
 
Notice finally that, both in the Appendix and in the following simplified
expressions, some factors resulting from the elementary partonic
cross sections have canceled out in the numerator and the
denominator of the polarization. When averaging the polarization over
some kinematical range, those factors must, of course, be taken into
account.

We consider separately different processes.
 
\goodbreak
\vskip6pt
\nd
{\bf a) $\ell^{\mp} \, p\to\ell^{\mp} \, \Lup \,X$}
\nobreak

This case is of interest for several experimental setups, {\it e.g.}\
HERMES, H1 and ZEUS at DESY, COMPASS at CERN, E665 at SLAC.
One gets
\bea
P_{_T}^{\,\Lambda}(x,y,z_h,p_{_T}) &=&
\frac{\sum_q\,e_q^2\,f_{q/p}(x)\,\left[\,d\hat\sigma^{\ell q}/
dy\,\right]\,\Delta^{\!N}\!D_{\Lup\!/q}(z_h,p_{_T})}
{\sum_q\,e_q^2\,f_{q/p}(x)\,\left[\,d\hat\sigma^{\ell q}/
dy\,\right]\,\hat{D}_{\Lambda/q}(z_h,p_{_T})}\nonumber\\
&\simeq& \frac{(4u+d)\,\Delta^{\!N}\!D_{\Lup\!/u}
+ s\,\Delta^{\!N}\!D_{\Lup\!/s}}
{(4u+d)\,\hat{D}_{\Lambda/u}+s\,\hat{D}_{\Lambda/s}} \>,
\label{ncl}
\eea
where in the second line, as discussed above, we have neglected nonleading, 
doubly suppressed, contributions, originating from sea quarks both in 
distribution and fragmentation functions; we have switched to the notation 
$f_{q/p}(x)\to q(x)$ and have assumed isospin symmetry to hold, that is
$\hat{D}_{\Lambda/d}\equiv\hat{D}_{\Lambda/u}$, and similarly for the
polarizing fragmentation functions.

The full expressions of the Appendix show that a similar simplified
expression holds also for the $\ell^{\mp}\,p\to\ell^{\mp}\,
\bar\Lambda^{\uparrow}\,X$ processes, with the exchange
$q(x) \leftrightarrow \bar{q}(x)$ (which may imply different
cancellation effects between the $u$, $d$ and $s$ polarizing FF
in the numerator) and with some additional terms in the denominator 
coming from nonleading quark contributions to the unpolarized FF (which in
turn may imply a stronger overall suppression factor as compared to the 
$\Lambda$ case).

\vskip6pt
\nd
{\bf b) $\nu\,p\to\nu\,\Lambda^\uparrow\,X$}

This process is of interest for the planned neutrino factories \cite{fmr,yb}, 
and is currently under investigation by the NOMAD Collaboration at CERN. Since 
there is no measurement of the final lepton here, care must be taken in the
definition of the appropriate kinematical range, which must exclude
regions where our perturbative approach is unreliable
(low $Q^2$ and $W^2$ regimes, {\it etc.}). The full expression of 
$P_{_T}^{\,\Lambda}$ is cumbersome; we give here the much simpler 
expression found when nonleading quark contributions are neglected:
\bea
P_{_T}^{\,\Lambda}(x,y,z_h,p_{_T}) &\simeq&
\frac{\left[\,u\,(1-8C)+d\,(1-4C)\,\right]\,
\Delta^{\!N}\!D_{\Lup\!/u}+
s\,(1-4C)\,\Delta^{\!N}\!D_{\Lup\!/s}}
{\left[\,u\,(1-8C)+d\,(1-4C)\,\right]\,\hat{D}_{\Lambda/u}+
s\,(1-4C)\,\hat{D}_{\Lambda/s}}\nonumber\\
&=& \frac{(Ku+d)\,\Delta^{\!N}\!D_{\Lup\!/u}+
s\,\Delta^{\!N}\!D_{\Lup\!/s}}
{(Ku+d)\,\hat{D}_{\Lambda/u}+s\,\hat{D}_{\Lambda/s}} \>,
\label{ncnu}
\eea
where $C=\sin^2\theta_{_W}/3 \simeq 0.077$, $K=(1-8C)/(1-4C) \simeq 0.55$ 
and terms quadratic in $C$, which only introduce a weak dependence on 
$y$ into $K$, have also been neglected.

An analogous expression holds for $\bar\nu\,p\to\bar\nu\,\Lup\,X$ 
processes, with a factor $K$ ranging now between 0.78 and 4 for $y=0$ and 
$y=1$ respectively.

Again, similar expressions hold also for $\nu\,p\to\nu\,\bar\Lup\,X$ 
and $\bar\nu\,p\to\bar\nu\,\bar\Lup\,X$ cases, with the exchange
$q(x) \leftrightarrow \bar{q}(x)$
and with some additional terms in the denominator coming from nonleading 
quark contributions to the unpolarized FF, which must be taken into account.

\vskip6pt
\nd
{\bf c) $\nu\,p\to\ell^-\,\Lambda^\uparrow\,X$}

This case is again of interest for neutrino factories and for the NOMAD 
experiment, which quite recently has published results for
$\Lambda$ and $\bar{\Lambda}$ polarization \cite{nomad00,nomad01}. 
One finds
\be
P_{_T}^{\,\Lambda}(x,y,z_h,p_{_T})=\frac{(d+R\,s)\,
\Delta^{\!N}\!D_{\Lup\!/u}+\bar{u}\,
(\Delta^{\!N}\!D_{\Lup\!/\bar{d}}+
R\,\Delta^{\!N}\!D_{\Lup\!/\bar{s}})\,
(1-y)^2}{(d+R\,s)\,\hat{D}_{\Lambda/u}+\bar{u}\,
(\hat{D}_{\Lambda/\bar{d}}+R\,\hat{D}_{\Lambda/\bar{s}})\,
(1-y)^2} \>,
\label{ccnu}
\ee
where $R=\tan^2\theta_C\simeq0.056$. Notice that
neglecting sea-quark contributions in the partonic distributions
and in the FF leads to a remarkably simple expression, which gives
direct access to the polarizing FF for $u$ quarks: 
\be
P_{_T}^{\,\Lambda}(x,y,z_h,p_{_T})\simeq
\frac{\Delta^{\!N}\!D_{\Lup\!/u}}
{\hat{D}_{\Lambda/u}} \>\cdot
\label{ccnusim}
\ee

The same expression is true for the case
$\ell^+\,p\to\bar\nu\,\Lup\,X$, which may be of interest for the H1 and 
ZEUS experiments at HERA, for COMPASS at CERN and E665 at SLAC.

Similar expressions also hold for the processes
$\bar\nu\,p \to \ell^+\,\bar\Lambda^{\uparrow}\,X$ and
$\ell^-\,p\to\nu\,\bar\Lambda^{\uparrow}\,X$ again with some
additional terms (different in the two cases) in the denominators coming 
from the nonleading quark contributions to the unpolarized FF. 

\vskip6pt
\nd
{\bf d) $\bar{\nu}\,p\to\ell^+\,\Lambda^\uparrow\,X$}

This case is very close to the previous one, with obvious
modifications:
\be
P_{_T}^{\,\Lambda}(x,y,z_h,p_{_T})=
 \frac{(1-y)^2\,u\,(\Delta^{\!N}\!D_{\Lup\!/d}+
 R\,\Delta^{\!N}\!D_{\Lup\!/s})+
 (\bar{d}+R\,\bar{s})\,\Delta^{\!N}\!D_{\Lup\!/\bar{u}}}
 {(1-y)^2\,u\,(\hat{D}_{\Lambda/d}+R\,\hat{D}_{\Lambda/s})+
 (\bar{d}+R\,\bar{s})\,\hat{D}_{\Lambda/\bar{u}}} \>\cdot
\label{ccnub}
\ee
Again, when sea-quark contributions are neglected and isospin
symmetry is invoked, we find the very simple expression

\be
P_{_T}^{\,\Lambda}(x,y,z_h,p_{_T})\simeq
\frac{\Delta^{\!N}\!D_{\Lup\!/u}+
R\,\Delta^{\!N}\!D_{\Lup\!/s}}
{\hat{D}_{\Lambda/u}+R\,\hat{D}_{\Lambda/s}} \>\cdot
\label{ccnubsim}
\ee

The same expression holds for the $\ell^-\,p\to\nu\,\Lup\,X$ case.

Once more, similar expressions also hold for the processes
$\nu\,p\to\ell^-\,\bar\Lambda^{\uparrow}\,X$ and
$\ell^+\,p\to\bar\nu\,\bar\Lambda^{\uparrow}\,X$ apart from some
additional terms (again different in the two cases) in the denominators
coming from the nonleading quark contributions to the unpolarized FF. 

\vskip6pt
The above results, Eqs.\ (\ref{ncl})-(\ref{ccnubsim}), relate measurable
polarizations to different combinations of (known) distribution functions,
(less known) unpolarized fragmentation functions and (unknown) 
polarizing fragmentation functions; the different terms have relative 
coefficients which depend on the dynamics of the elementary partonic 
process and/or on the relevance of $s$ quark contributions
in the partonic distribution functions (the latter can be modulated
according to the specific $x$ region explored). 

This large variety of possibilities gives a good opportunity to investigate 
and test the relevant properties of the unpolarized and polarizing $\Lambda$
FF, by measuring the hyperon transverse polarization. In some special 
cases, Eq.\ (\ref{ccnusim}), experiment offers direct information on these 
new functions. 

\vspace{18pt}
\goodbreak
\nd
{\bf 5. Numerical estimates}
\nobreak
\vspace{6pt}
\nobreak

We derive some numerical results for $P_{_T}^{\,\Lambda}$ and 
$P_{_T}^{\,\bar\Lambda}$, according to a few plausible scenarios for 
the fragmentation functions given in Eqs.\ (\ref{dfin}) and (\ref{dedfin}).
In the numerical calculations we have utilized the full expressions of the 
polarization given in the Appendix, with the only simplification 
$\Delta^{\!N}\!D_{\Lup\!/\bar q} = 0$: these full expressions differ 
from the approximated ones (\ref{ncl})-(\ref{ccnubsim}) by extra terms
in the denominators which, as we have explicitly checked, give 
negligible contributions to the numerical results presented here.

Let us consider the parameters $\alpha, \beta, N_q$ and $r$ appearing in 
Eqs. (\ref{dedfin}) and (\ref{pardel}), and their possible values.
Looking only qualitatively at the data on transverse $\Lambda$ polarization 
in hadronic reactions we expect the following general features. 
1) Negative contributions from up and down quarks 
($N_{u,d} < 0$) and positive from strange quarks ($N_s > 0$), in order to 
have $P^\Lambda <0$ and $P^{\bar\Lambda}\simeq 0$\footnote{This feature is 
similar to what is expected for the longitudinally polarized FF,
$\Delta D_{\Lambda/q}(z)$, in the well-known
Burkardt-Jaffe model~\cite{buja93}.}; based on this one expects the 
$\Lambda$ polarization in semi-inclusive DIS, where the $u$ quark contribution
is enhanced by the charge, to be negative. 2) A polarizing FF peaked at 
large $z$ to explain the increase in magnitude of the polarization 
with $x_F$ at fixed $p_T$ (thus implying large values of $\alpha$, while 
$\beta \simeq O(1)$). 3) A Gaussian shape similar for unpolarized and 
polarizing FF to explain the large values of the polarization (which means 
$r\simeq O(1)$). 

We then fix $\beta = 1$, $\alpha = 6$, $r = 0.7$ and follow 
Ref.~\cite{abm95} using
\be 
\sqrt{\langle k_\perp^{\!\ 2}(z)\rangle} = 0.61\, z^{0.27}(1-z)^{0.2}
\> {\rm GeV}/c \>, \label{ktpion}
\ee
as given by fitting the pion $k_\perp$ distribution inside jets at LEP 
energies.

We adopt for the unpolarized, $\bfk_\perp$-$\,$integrated, $\Lambda$ FF, 
the $SU(3)$ symmetric parameterizations of Ref.~\cite{boros2} (BLT).
Other interesting sets of $SU(3)$ symmetric fragmentation functions were
obtained in Ref.\ \cite{fsv}. However, these sets do not separate between
$\Lambda$ and $\bar\Lambda$ and we cannot use them here, where we compute
separately $P_{_T}^\Lambda$ and $P_{_T}^{\bar\Lambda}$; this was not a 
problem in Ref.\ \cite{abdm01} as, in the kinematical regions considered 
there for $p\,p \to \Lup \,X$, it turned out that $P_{_T}^\Lambda
\simeq P_{_T}^{\Lambda + \bar\Lambda}$. We have checked that,
at least for HERMES and NOMAD kinematics (for which
$d\sigma^\Lambda \gg d\sigma^{\bar\Lambda}$), the $\Lambda + \bar\Lambda$
fragmentation functions of Ref.\ \cite{fsv} give here results close to 
those obtained with the symmetric BLT set. 

$SU(3)$ breaking unpolarized FF are also available in the literature
\cite{indu}; however, we do not aim here at discriminating among 
different FF, and we do not consider all possible cases. Such an analysis 
can only be attempted when much more experimental information will be
available; at that stage the full analytical formalism discussed here
can be used to learn about quark fragmentation processes into $\Lambda's$.   

We look at qualitative differences of the results by considering for the 
polarizing FF two different models (keeping the BLT set of unpolarized FF):

\begin{enumerate}
\item
a scenario with almost the same weight for up and strange quarks, 
with $N_u=N_d=-0.8$ and $N_s=1$;
\item
a scenario similar to the model of Burkardt and Jaffe \cite{buja93}  
for the longitudinally polarized FF $\Delta D_{\Lambda/q}$, with
$N_u=N_d=-0.3$ and $N_s=1$. 
\end{enumerate}

We have adopted the MRST99 \cite{mrst99} parameterization for the
unpolarized proton distribution functions.

We consider kinematical configurations typical of running experiments
(HERMES at DESY, NOMAD at CERN, E655 at SLAC) that are presently measuring 
transverse $\Lambda$ polarization or plan to do it in the near future.
The main kinematical cuts considered are: 

\nd
1) HERMES: $0.023 < x < 0.4$,
$y < 0.85$, $1 < Q^2 < 10$ (GeV/$c)^2$, $E_\Lambda > 4.5$ GeV; 

\nd
2) NOMAD:
$\langle x \rangle = 0.22$, $\langle y \rangle = 0.48$, $\langle Q^2 \rangle
= 9$ (GeV/$c)^2$; 

\nd
3) E665: $10^{-3} < x < 10^{-1}$, $0.1 < y < 0.8$,
$1.0 < Q^2 < 2.5$ (GeV/$c)^2$, $E_\Lambda > 4$ GeV.

Since the $Q^2$ evolution of the polarizing FF is not under control
at present, and the HERMES and NOMAD experiments involve a
relatively limited range of $Q^2$ values, in our numerical
calculations we have adopted a fixed scale, $Q^2=2$ (GeV/$c)^2$.
The unpolarized FF of Ref.~\cite{boros2} are given at a very
low $Q^2$ value, and no evolution program is available at present from the
authors; we have then performed the proper evolution to the adopted scale
by using the evolution codes of Miyama and Kumano \cite{kumano}.       
       
Our results are shown in Figs.\ \ref{pla} and \ref{plb}. Fig.\ \ref{pla} 
shows 
$P_{_T}^{\Lambda}$ as a function of $z_h$, after $p_{_T}$ average,
for all processes {\bf a)-d)} considered in the previous Section and for 
kinematical configurations typical of the corresponding relevant experiments 
(as indicated in the legends); the two plots (left and right) correspond,
respectively, to scenario 1 and 2 of the polarizing FF.

The polarization is in general large in magnitude and negative:
contributions from $\Delta^{\!N}\!D_{\Lup\!/s}$ are always suppressed, 
either by the $s$ quark distribution [via factors like $s/(K\,u + d)]$ 
or by the Standard Model factor $R$, see Eqs.\ (\ref{ncl})-(\ref{ccnubsim}). 
Thus, the strange quark contribution is suppressed, unless one uses a 
$SU(3)$ asymmetric FF set for which $|\Delta^{\!N}\!D_{\Lup\!/s}| \gg 
|\Delta^{\!N}\!D_{\Lup\!/u,d}|$. 
This is reflected by the fact that in Fig.\ \ref{pla} 
all processes have similar polarizations, approximately given by 
the $p_{_T}$-averaged ratio $\Delta^{\!N}\!D_{\Lup\!/u}/\hat{D}_{\Lambda/u}$. 
The polarization is smaller in the right plot (scenario 2) simply because
$|\Delta^{\!N}\!D_{\Lup\!/u}|$ is smaller.

Fig.~\ref{plb} shows the corresponding results for the case of 
$\bar\Lambda$ SIDIS production; here the effect of
cancellations between up (down) and strange contributions 
is more significant: notice, for instance, how 
$P_{_T}(\nu\,p\to\nu\,\bar\Lambda^{\uparrow}\,X)$ is suppressed 
compared to the analogous process for $\Lambda$. 

\vspace{18pt}
\goodbreak
\nd
{\bf 6. Conclusions}
\label{concl}
\nobreak
\vspace{6pt}
\nobreak

Single spin effects, which appear to be suppressed in leading twist collinear 
applications of pQCD factorization theorems, may instead reveal new 
interesting aspects of nonperturbative QCD. In Ref.\ \cite{abdm01} and 
in the present paper we have considered the long-standing problem of the
transverse polarization of hyperons produced from {\it unpolarized}
initial nucleons. Our approach is based on the use of a QCD factorization 
scheme, generalized to include intrinsic $\bfk_\perp$ in the
fragmentation process: this allows to introduce new spin dependences
in the fragmentation functions of unpolarized quarks, which may cause the 
observed polarization of the produced hyperons. 

These new functions, the polarizing FF, are supposed to describe universal 
features of the hadronization process, which is factorized in a similar
way as the usual $\bfk_\perp$-$\,$integrated fragmentation functions.
If correct, this idea should allow for a consistent phenomenological 
description of hyperon polarization in different processes: once the 
polarizing FF are extracted from some sets of data, their use in other 
processes should yield genuine predictions. With present experimental 
knowledge this does not appear to be feasible, unfortunately.

In Ref.\ \cite{abdm01} we have parameterized in a simple way the new functions
and, by fitting all data on $p\,p \to \Lup\,X$ and $p\,Be \to \Lup\,X$ with
$p_{_T} > 1$ GeV$/c$, we have obtained explicit expressions for the polarizing
FF; however, as the study of the unpolarized cross section reveals, 
these data seem not to be in the appropriate kinematical region yet, 
so that the extracted polarizing FF cannot be safely used to give true 
predictions. Data which unambiguously originate from kinematical regions 
where pQCD and factorization theorems can be applied, with a clear signature 
of current quark fragmentation, is still awaited. Therefore, the previously 
obtained expressions have been utilized here only in order to construct 
different scenarios for the polarizing FF and, hence, for the transverse 
$\Lambda$ polarization in SIDIS processes, 
$\ell \, p \to \ell' \, \Lup \, X$. In this
respect our present results should not be interpreted as predictions. 
But our results do display some interesting qualitative features that seem to
be generic and are 
worth testing, for instance the $\bar \Lambda$ polarization is in most cases
comparable in size to the $\Lambda$ polarization. 

We would like to stress once more that our approach is describing
a different kinematical region compared to the models based on target
fragmentation, which aim at explaining $\Lambda$ polarization produced in 
low $p_T$ data in $p$-$p$ collisions and at $x_{_F}<0$ data in semi-inclusive
DIS. In contrast, our present investigation applies to the production of 
$\Lambda$'s in semi-inclusive DIS for $x_{_F}>0$. Relevant 
data are expected to be available soon and a comparison with our analytical
results will then be very useful in determining the polarizing FF.

\vskip 18pt
\goodbreak
\nd
{\bf Acknowledgements}
\vskip 6pt
We would like to thank A. Efremov, J. Soffer and W. Vogelsang for many
useful discussions and for pointing out to us data on unpolarized 
cross sections for $\Lambda$ production and difficulties in their 
understanding within pQCD factorization scheme. 
D.B. thanks the RIKEN-BNL Research Center where most of his work on this 
topic has been performed. At present, the research of D.B. has been made 
possible by financial support from the Royal Netherlands Academy of Arts and 
Sciences. U.D. and F.M. thank COFINANZIAMENTO MURST-PRIN for partial support.

\vspace{18pt}
\goodbreak
\nd
{\bf Appendix}
\nobreak
\vspace{6pt}
\nobreak

We present here the full analytical expressions for the transverse 
polarization, $P_{_T}$, of $\Lambda$ and $\bar\Lambda$ hyperons produced
in semi-inclusive DIS, at leading order in the coupling constants and in 
a $1/Q$ expansion; we neglect, for $(\ell^\pm,\ell^\pm)$ processes, 
electroweak interference effects. Our kinematical configuration is
defined in the virtual boson-target proton c.m. reference frame, 
explained in detail in the text (see Fig. 1). The polarization is a 
function of the usual DIS variables 
$x=Q^2/2p\cdot{q}$, $y=p \cdot q/p \cdot \ell$ and 
$z_h = p\cdot p_h/p \cdot q$,
where $\ell$, $p$, $q$, $p_h$ are the 4-momenta of the initial
lepton, the proton, the virtual boson and the observed hadron
respectively. It depends also on $p_{_T}$ ($\equiv k_\perp$ in our 
configuration), through the FF.
For simplicity, all these dependences are not indicated
in the following equations. Similarly, the $Q^2$ evolution is also implicit.
We use the notation $q=u,d,s$ for the partonic distributions and
in all cases we refer to $\Lambda$ fragmentation functions,
since $(\Delta^{\!N})D_{\bar\Lambda/q,\bar{q}}\equiv
(\Delta^{\!N})D_{\Lambda/\bar{q},q}$. The factors
$(1-y)^2$ result from the elementary partonic cross sections involving
leptons and quarks (or antiquarks) with opposite helicities.      
In all expressions we always put first terms which
contain the dominant fragmentation functions. $R$ and $C$ are typical 
Standard Model factors, $R=\tan^2\theta_C \simeq 0.056$, and
$C=\sin^2\theta_{_W}/3\simeq 0.077$.

An inspection of the full set of equations shows that there are
simple ways of relating the various processes:\\
-- The polarizations for $\Lambda$ and $\bar\Lambda$
involving the same leptons are simply connected by exchanging
$(\Delta^{\!N})D_{\Lambda/q}$ with $(\Delta^{\!N})D_{\Lambda/\bar{q}}$;\\
-- Both for $\Lambda$ and $\bar\Lambda$, the polarizations involving
respectively $(\nu,\ell^-)$, $(\nu,\nu)$, $(\ell^-,\nu)$ and 
$(\bar\nu,\ell^+)$, $(\bar\nu,\bar\nu)$, $(\ell^+,\bar\nu)$ processes are 
related by exchanging $q$ with $\bar{q}$  and $(\Delta^{\!N})D_{\Lambda/q}$
with $(\Delta^{\!N})D_{\Lambda/\bar{q}}$;\\
--  Both for $\Lambda$ and $\bar\Lambda$, the polarizations involving
``crossed'' processes like $(\nu,\ell^-)$ and $(\ell^+,\bar\nu)$
are connected by exchanging the position of the $(1-y)^2$ factors in
the terms involving quark and antiquark distributions.
This can be easily understood since one process and its ``crossed'' 
one both correspond to the exchange of the same $W$ boson, but with a negative
and positive helicity lepton respectively.

\vspace{12pt}

\noindent {\bf a) \mbox{\boldmath$\Lambda$} polarization}

\vspace{6pt}

{\bf a1) $\nu\,p\to\ell^-\,\Lambda^{\uparrow}\,X$}

\be
 P_{_T} = \frac{(d+R\,s)\,\Delta^{\!N}\!D_{\Lup\!/u}+
 (1-y)^2\,\bar{u}\,(\Delta^{\!N}\!D_{\Lup\!/\bar{d}}+
 R\,\Delta^{\!N}\!D_{\Lup\!/\bar{s}})}
 {(d+R\,s)\,\hat{D}_{\Lambda/u}+(1-y)^2\,\bar{u}\,
 (\hat{D}_{\Lambda/\bar{d}}+R\,\hat{D}_{\Lambda/\bar{s}})}
\label{a-nulm}
\ee

{\bf a2) $\bar\nu\,p\to\ell^+\,\Lambda^{\uparrow}\,X$}  

\be
 P_{_T} =
 \frac{(1-y)^2\,u\,(\Delta^{\!N}\!D_{\Lup\!/d}+
 R\,\Delta^{\!N}\!D_{\Lup\!/s})+
 (\bar{d}+R\,\bar{s})\,\Delta^{\!N}\!D_{\Lup\!/\bar{u}}}
 {(1-y)^2\,u\,(\hat{D}_{\Lambda/d}+R\,\hat{D}_{\Lambda/s})+
 (\bar{d}+R\,\bar{s})\,\hat{D}_{\Lambda/\bar{u}}}
\label{a-nblp}
\ee

{\bf a3) $\ell^-\,p\to\nu\,\Lambda^{\uparrow}\,X$}

\be
 P_{_T} =
 \frac{u\,(\Delta^{\!N}\!D_{\Lup\!/d}+ R\,\Delta^{\!N}\!D_{\Lup\!/s})+
 (1-y)^2\,(\bar{d}+R\,\bar{s})\,\Delta^{\!N}\!D_{\Lup\!/\bar{u}}}
 {u\,(\hat{D}_{\Lambda/d}+R\,\hat{D}_{\Lambda/s})+
 (1-y)^2\,(\bar{d}+R\,\bar{s})\,\hat{D}_{\Lambda/\bar{u}}}
 \label{a-lmnu}
\ee

{\bf a4) $\ell^+\,p\to\bar\nu\,\Lambda^{\uparrow}\,X$}

\be
 P_{_T} =
 \frac{(1-y)^2\,(d+R\,s)\,\Delta^{\!N}\!D_{\Lup\!/u}+
 \bar{u}\,(\Delta^{\!N}\!D_{\Lup\!/\bar{d}}+R\,
 \Delta^{\!N}\!D_{\Lup\!/\bar{s}})}
 {(1-y)^2\,(d+R\,s)\,\hat{D}_{\Lambda/u}+
 \bar{u}\,(\hat{D}_{\Lambda/\bar{d}}+
 R\,\hat{D}_{\Lambda/\bar{s}})}
 \label{a-lpnb}
\ee
 
{\bf a5) $\ell^{\mp}\,p\to\ell^{\mp}\,\Lambda^{\uparrow}\,X$}

\be
 P_{_T} =
 \frac{4\,u\,\Delta^{\!N}\!D_{\Lup\!/u}+d\,\Delta^{\!N}\!D_{\Lup\!/d}+
 s\,\Delta^{\!N}\!D_{\Lup\!/s}+4\,\bar{u}\,\Delta^{\!N}\!D_{\Lup\!/\bar{u}}+
 \bar{d}\,\Delta^{\!N}\!D_{\Lup\!/\bar{d}}+
 \bar{s}\,\Delta^{\!N}\!D_{\Lup\!/\bar{s}}}
 {4\,u\,\hat{D}_{\Lambda/u}+d\,\hat{D}_{\Lambda/d}+s\,\hat{D}_{\Lambda/s}+
 4\,\bar{u}\,\hat{D}_{\Lambda/\bar{u}}+\bar{d}\,\hat{D}_{\Lambda/\bar{d}}+
 \bar{s}\,\hat{D}_{\Lambda/\bar{s}}}
\label{a-lmlm}
\ee

{\bf a6) $\nu\,p\to\nu\,\Lambda^{\uparrow}\,X$}

\bea
\vspace*{-8pt}
{\rm Num}(P_{_T})
 &=& \left[\,(1-4C)^2+(1-y)^2\,16C^2\,\right]\,
 u\,\Delta^{\!N}\!D_{\Lup\!/u}\nonumber\\
 &+&\left[\,(1-2C)^2+(1-y)^2\,4C^2\,\right]\,
 (d\,\Delta^{\!N}\!D_{\Lup\!/d}+s\,\Delta^{\!N}\!D_{\Lup\!/s})\nonumber\\
 &+&\left[\,(1-y)^2\,(1-4C)^2+16C^2\,\right]\,
 \bar{u}\,\Delta^{\!N}\!D_{\Lup\!/\bar{u}}\nonumber\\
 &+&\left[\,(1-y)^2\,(1-2C)^2+4C^2\,\right]\,
 (\bar{d}\,\Delta^{\!N}\!D_{\Lup\!/\bar{d}}+
 \bar{s}\,\Delta^{\!N}\!D_{\Lup\!/\bar{s}})\nonumber\\
 \mbox{}\hspace{-1.9truecm} {\rm Den}(P_{_T})&:&
 \Delta^{\!N}\!D_{\Lup\!/q} \to \hat{D}_{\Lambda/q}
\label{a-nunu}
\eea

{\bf a7) $\bar\nu\,p\to\bar\nu\,\Lambda^{\uparrow}\,X$}

\bea
\vspace*{-8pt}
 {\rm Num}(P_{_T}) &=&
 \left[\,(1-y)^2\,(1-4C)^2+16C^2\,\right]\,
 u\,\Delta^{\!N}\!D_{\Lup\!/u}\nonumber\\
 &+&\left[\,(1-y)^2\,(1-2C)^2+4C^2\,\right]\,
 (d\,\Delta^{\!N}\!D_{\Lup\!/d}+s\,\Delta^{\!N}\!D_{\Lup\!/s})\nonumber\\
 &+&\left[\,(1-4C)^2+(1-y)^2\,16C^2\,\right]\,
 \bar{u}\,\Delta^{\!N}\!D_{\Lup\!/\bar{u}}\nonumber\\
 &+&\left[\,(1-2C)^2+(1-y)^2\,4C^2\,\right]\,
 (\bar{d}\,\Delta^{\!N}\!D_{\Lup\!/\bar{d}}+
 \bar{s}\,\Delta^{\!N}\!D_{\Lup\!/\bar{s}})\nonumber\\
 \mbox{}\hspace{-1.9truecm} {\rm Den}(P_{_T})&:&
 \Delta^{\!N}\!D_{\Lup\!/q} \to \hat{D}_{\Lambda/q} \>.
\label{a-nbnb}
\eea

Notice that often one can safely neglect terms involving 
sea (or nonleading) quark contributions both in the partonic
distributions and fragmentation functions, which in some
cases leads to very simple expressions for the polarization.
Some examples are presented and discussed in Section 4.
However, in the cases involving $\bar\nu$'s, because of the terms
$(1-y)^2$, these approximate expressions should be taken with some
care in the region $y\simeq 1$. Further simplifications
result when assuming isospin invariance,
$(\Delta^{\!N})D_{\Lambda/u}\equiv(\Delta^{\!N})D_{\Lambda/d}$.

\vspace{12pt}

\noindent {\bf b) \mbox{\boldmath$\bar\Lambda$} polarization}

\vspace{6pt}

{\bf b1) $\nu\,p\to\ell^-\,\bar\Lambda^{\uparrow}\,X$}

\be
 P_{_T} = 
 \frac{(1-y)^2\,\bar{u}\,(\Delta^{\!N}\!D_{\Lup\!/d}+
 R\,\Delta^{\!N}\!D_{\Lup\!/s})+(d+R\,s)\,\Delta^{\!N}\!D_{\Lup\!/\bar{u}}}
 {(1-y)^2\,\bar{u}\,(\hat{D}_{\Lambda/d}+\hat{D}_{\Lambda/s})+
 (d+R\,s)\,\hat{D}_{\Lambda/\bar{u}}}
\label{b-nulm}
\ee

{\bf b2) $\bar\nu\,p\to\ell^+\,\bar\Lambda^{\uparrow}\,X$}

\be
 P_{_T} =
 \frac{(\bar{d}+R\,\bar{s})\,\Delta^{\!N}\!D_{\Lup\!/u}+
 (1-y)^2\,u\,(\Delta^{\!N}\!D_{\Lup\!/\bar{d}}+
 R\,\Delta^{\!N}\!D_{\Lup\!/\bar{s}})}
 {(\bar{d}+R\,\bar{s})\,\hat{D}_{\Lambda/u}+(1-y)^2\,u\,
 (\hat{D}_{\Lambda/\bar{d}}+R\,\hat{D}_{\Lambda/\bar{s}})}
\label{b-nblp}
\ee

{\bf b3) $\ell^-\,p\to\nu\,\bar\Lambda^{\uparrow}\,X$}

\be
 P_{_T} =
 \frac{(1-y)^2\,(\bar{d}+R\,\bar{s})\,\Delta^{\!N}\!D_{\Lup\!/u}+
 u\,(\Delta^{\!N}\!D_{\Lup\!/\bar{d}}+R\,\Delta^{\!N}\!D_{\Lup\!/\bar{s}})}
 {(1-y)^2\,(\bar{d}+R\,\bar{s})\,\hat{D}_{\Lambda/u}+
 u\,(\hat{D}_{\Lambda/\bar{d}}+R\,\hat{D}_{\Lambda/\bar{s}})}
\label{b-lmnu}
\ee

{\bf b4) $\ell^+\,p\to\bar\nu\,\bar\Lambda^{\uparrow}\,X$}

\be
 P_{_T} =
 \frac{\bar{u}\,(\Delta^{\!N}\!D_{\Lup\!/d}+R\,\Delta^{\!N}\!D_{\Lup\!/s})+
 (1-y)^2\,(d+R\,s)\,\Delta^{\!N}\!D_{\Lup\!/\bar{u}}}
 {\bar{u}\,(\hat{D}_{\Lambda/d}+R\,\hat{D}_{\Lambda/s})+
 (1-y)^2\,(d+R\,s)\,\hat{D}_{\Lambda/\bar{u}}}
\label{b-lpnb}
\ee

{\bf b5) $\ell^{\mp}\,p\to\ell^{\mp}\,\bar\Lambda^{\uparrow}\,X$}

\be
 P_{_T} =
 \frac{4\,\bar{u}\,\Delta^{\!N}\!D_{\Lup\!/u}+
 \bar{d}\,\Delta^{\!N}\!D_{\Lup\!/d}+\bar{s}\,\Delta^{\!N}\!D_{\Lup\!/s}
 +4\,u\,\Delta^{\!N}\!D_{\Lup\!/\bar{u}}+
 d\,\Delta^{\!N}\!D_{\Lup\!/\bar{d}}+s\,\Delta^{\!N}\!D_{\Lup\!/\bar{s}}}
 {4\,\bar{u}\,\hat{D}_{\Lambda/u}+\bar{d}\,\hat{D}_{\Lambda/d}+
 \bar{s}\,\hat{D}_{\Lambda/s}+4\,u\,\hat{D}_{\Lambda/\bar{u}}+
 d\,\hat{D}_{\Lambda/\bar{d}}+ s\,\hat{D}_{\Lambda/\bar{s}}}
\label{b-lmlm}
\ee

{\bf b6) $\nu\,p\to\nu\,\bar\Lambda^{\uparrow}\,X$}

\bea
\vspace*{-8pt}
{\rm Num}(P_{_T})
 &=& \left[\,(1-y)^2\,(1-4C)^2+16C^2\,\right]\,
 \bar{u}\,\Delta^{\!N}\!D_{\Lup\!/u}\nonumber\\
 &+&\left[\,(1-y)^2\,(1-2C)^2+4C^2\,\right]\,
 (\bar{d}\,\Delta^{\!N}\!D_{\Lup\!/d}+
 \bar{s}\,\Delta^{\!N}\!D_{\Lup\!/s})\nonumber\\
 &+&\left[\,(1-4C)^2+(1-y)^2\,16C^2\,\right]\,
 u\,\Delta^{\!N}\!D_{\Lup\!/\bar{u}}\nonumber\\
 &+&\left[\,(1-2C)^2+(1-y)^2\,4C^2\,\right]\,
 (d\,\Delta^{\!N}\!D_{\Lup\!/\bar{d}}+
 s\,\Delta^{\!N}\!D_{\Lup\!/\bar{s}})\nonumber\\
 \mbox{}\hspace{-1.9truecm} {\rm Den}(P_{_T})&:&
 \Delta^{\!N}\!D_{\Lup\!/q} \to \hat{D}_{\Lambda/q}
\label{b-nunu}
\eea

{\bf b7) $\bar\nu\,p\to\bar\nu\,\bar\Lambda^{\uparrow}\,X$}

\bea
\vspace*{-8pt}
 {\rm Num}(P_{_T}) &=&
 \left[\,(1-4C)^2+(1-y)^2\,16C^2\,\right]\,
 \bar{u}\,\Delta^{\!N}\!D_{\Lup\!/u}\nonumber\\
 &+&\left[\,(1-2C)^2+(1-y)^2\,4C^2\,\right]\,
 (\bar{d}\,\Delta^{\!N}\!D_{\Lup\!/d}+
 \bar{s}\,\Delta^{\!N}\!D_{\Lup\!/s})\nonumber\\
 &+&\left[\,(1-y)^2\,(1-4C)^2+16C^2\,\right]\,
 u\,\Delta^{\!N}\!D_{\Lup\!/\bar{u}}\nonumber\\
 &+&\left[\,(1-y)^2\,(1-2C)^2+4C^2\,\right]\,
 (d\,\Delta^{\!N}\!D_{\Lup\!/\bar{d}}+
 s\,\Delta^{\!N}\!D_{\Lup\!/\bar{s}})\nonumber\\
 \mbox{}\hspace{-1.9truecm} {\rm Den}(P_{_T})&:&
 \Delta^{\!N}\!D_{\Lup\!/q} \to \hat{D}_{\Lambda/q} \>.
\label{b-nbnb}
\eea

Notice that contrary to the case of $\Lambda$ particle production,
here the leading and nonleading quark contributions are mixed
between partonic distributions and fragmentation functions,
with terms of the type $q\,(\Delta^{\!N})D_{\Lambda/\bar{q}}$ and
$\bar{q}\,(\Delta^{\!N})D_{\Lambda/q}$. Which terms are dominating
depends on the kinematic range considered ($x$ and $z_h$ values).
Moreover, the $(1-y)^2$ factors can also be relevant, for large $y$
values. Therefore, it is not easy to find approximate expressions
for $P_{_T}$; in any case, their range of validity is limited to particular
kinematical configurations and has to be considered with care.

\clearpage
%

\begin{figure}[t]
\begin{center}
\hspace*{-0.2cm}
\mbox{~\epsfig{file=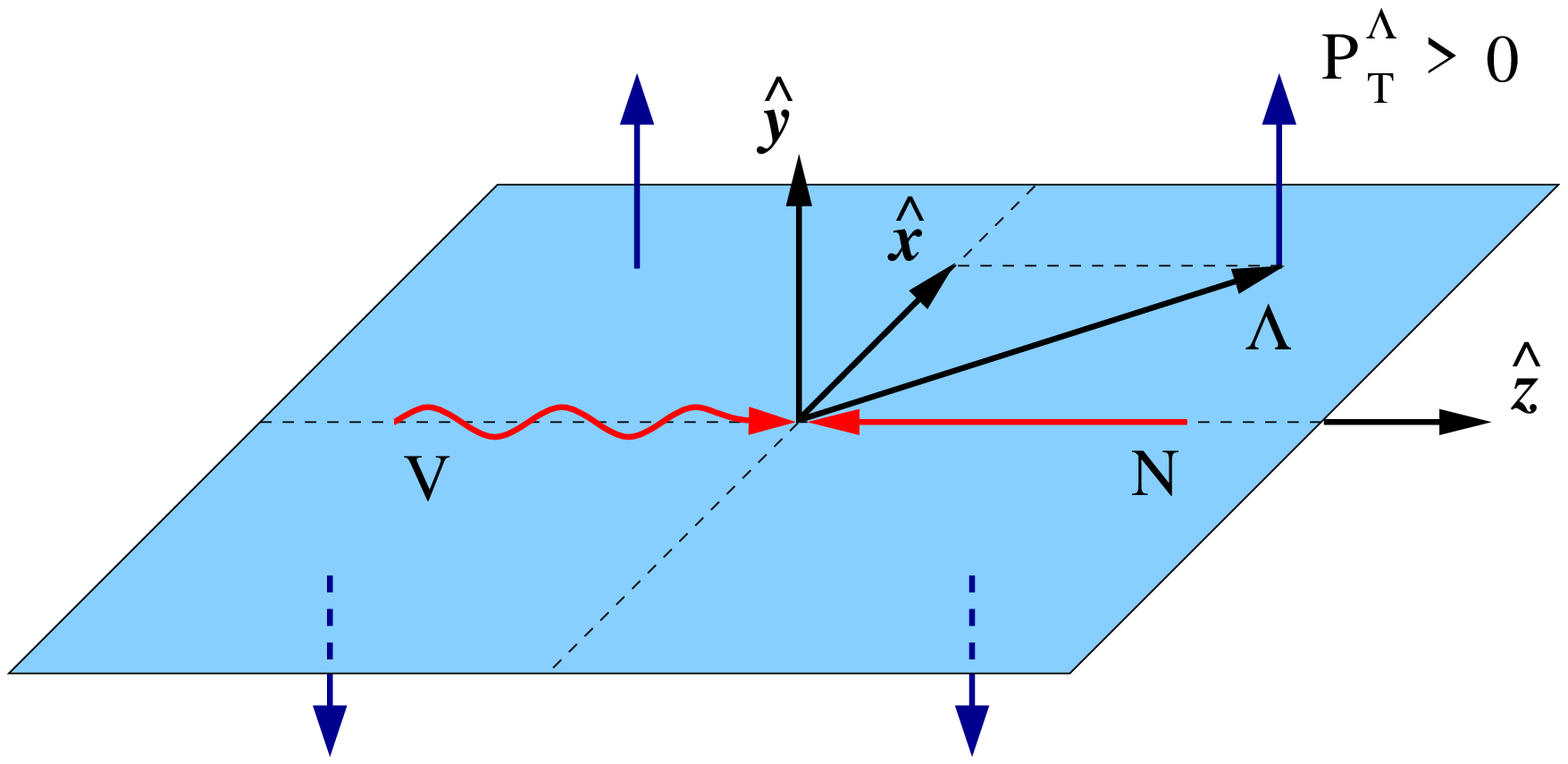,angle=0,width=12.5cm}}
\end{center}
\caption[f1]{\label{frame}
Kinematical representation of the semi-inclusive hadron production
process, as viewed in the virtual boson-target proton c.m. reference frame
(VN frame). The $\hat{z}$ axis is taken along the virtual boson
direction of motion; the $\hat{x}$ axis is along the direction of the
hadron transverse momentum, $\mbox{\boldmath$p$}_{_T}\equiv
\mbox{\boldmath$k$}_\perp$; the $\hat{y}$ axis is chosen to form a
right-handed frame. The direction of {\it positive}  transverse
(with respect to the production plane) polarization $\bfP_{_T}$,
in the four quadrants, is shown explicitly by the arrows. 
}
\end{figure}

\clearpage


\begin{figure}[t] 
\begin{center}
\hspace*{-0.6cm}
\mbox{~\epsfig{file=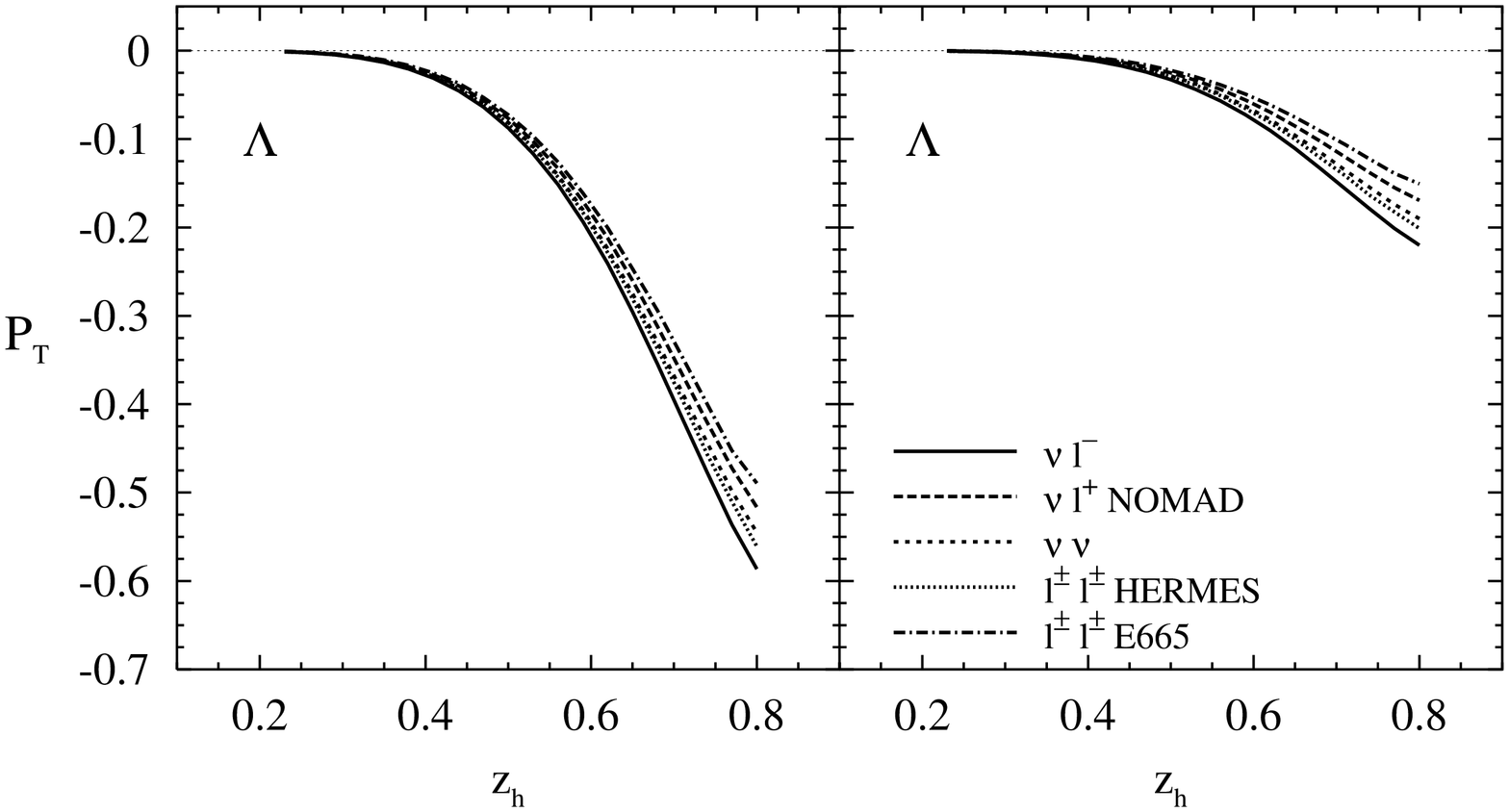,angle=-90,width=15.3cm}}
\end{center}
\caption[f2]
{\label{pla}
Transverse $\Lambda$ polarization, $P_{_T}^{\Lambda}$, {\it vs.} $z_h$ and
averaged over $p_{_T}$, for several SIDIS production processes, 
with scenario 1 (on the left) and scenario 2 (on the right) 
for the polarizing FF (see text for more details).
}
\end{figure}

\clearpage


\begin{figure}[t] 
\begin{center}
\hspace*{-0.6cm}
\mbox{~\epsfig{file=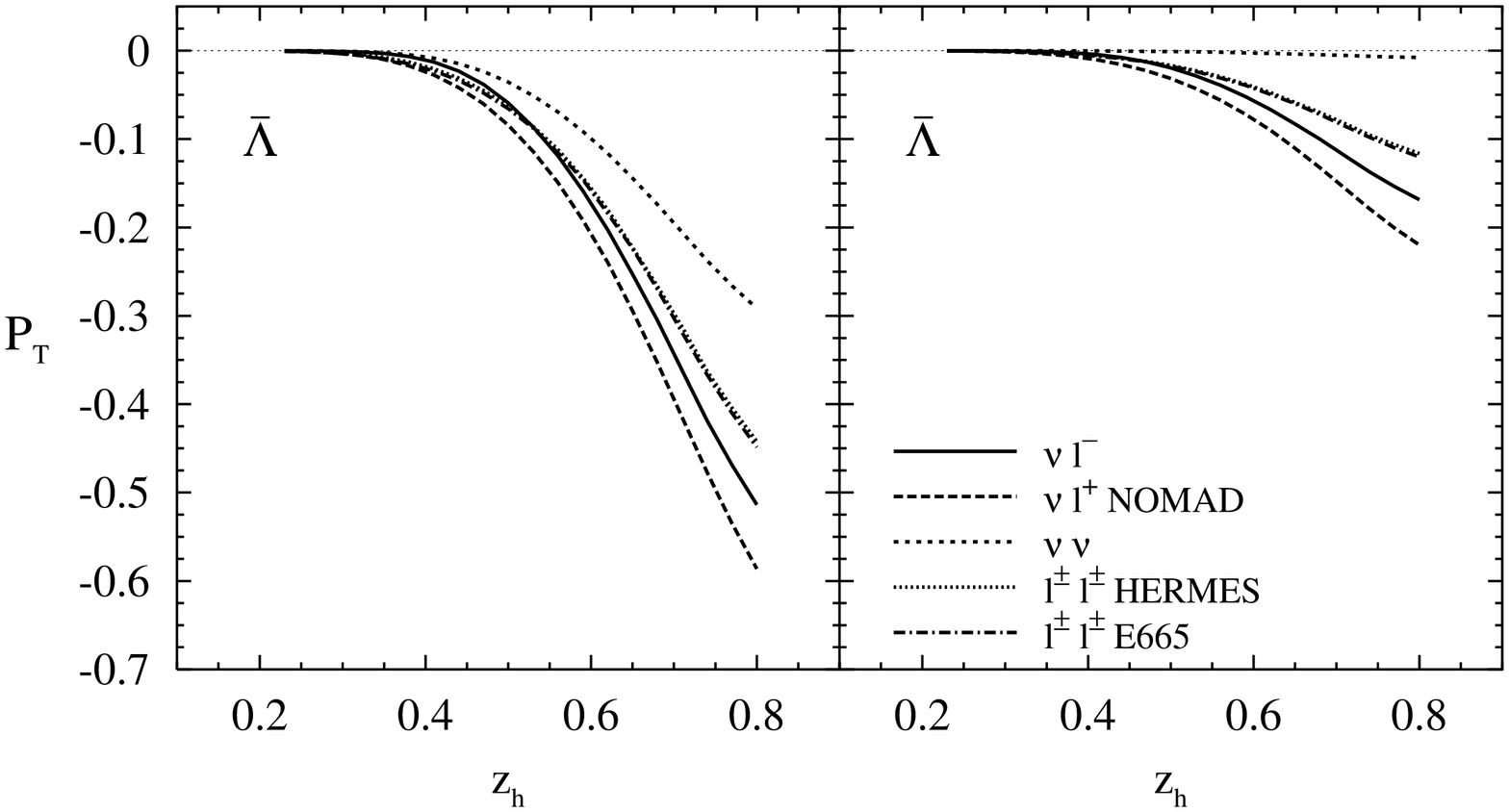,angle=-90,width=15.3cm}}
\end{center}
\caption[f2]
{\label{plb}
Transverse $\bar\Lambda$ polarization, $P_{_T}^{\bar\Lambda}$, {\it
vs.} $z_h$ and
averaged over $p_{_T}$, for several SIDIS production processes, 
with scenario 1 (on the left) and scenario 2 (on the right) 
for the polarizing FF (see text for more details).
}
\end{figure}
\end{document}